\documentclass[reprint,aps,prb,nofootinbib]{revtex4-2}

\usepackage{hyperref}
\usepackage[pdftex]{graphicx}
\usepackage{amsmath, amssymb, amsthm, braket,mathrsfs}
\usepackage[T1]{fontenc}
\usepackage{orcidlink}
\usepackage{bm}
\hypersetup{colorlinks=true}

\begin{document}


\preprint{APS/123-QED}

\title{Anyon Bound States and Hybrid Superconductivity}
\author{Paul~Leask\orcidlink{0000-0002-6012-0034}}
\email{palea@kth.se}
\affiliation{Department of Physics, KTH Royal Institute of Technology, 10691 Stockholm, Sweden}
\date{\today}

\begin{abstract}
The interactions of anyonic quasi-particles (vortices) in the Chern--Simons extension of the Ginzburg--Landau model is investigated and we show that it manifestly realizes a hybridization of type I/II superconductivity.
Through Gauss' law, each vortex simultaneously carries a flux quantum and a proportional Noether charge, thereby realizing an anyonic excitation.
The Chern--Simons coupling also modifies the screening structure of the gauge fields, producing complex-conjugate masses that yield a common penetration depth with an oscillatory phase.
This altered asymptotic behavior breaks the conventional type-I/type-II dichotomy of the Ginzburg--Landau model.
As a result, vortex anyons experience short-range repulsion and long-range attraction, enabling the formation of separated multi-vortex bound states with non-monotonic interaction energy.
\end{abstract}

\maketitle



\emph{Introduction.}
Topological phases of matter in two spatial dimensions support exotic excitations known as anyons: quasi-particles with fractional statistics interpolating between bosons and fermions.
Anyons emerge in diverse condensed matter systems, including the fractional quantum Hall effect (FQHE), where electric charge-magnetic flux attachment generates excitations of fractional charge \cite{ZHK_1989,Zhang_1992,Mulligan_2010}, and in spinor Bose--Einstein condensates where vortex braiding realizes anyonic exchange statistics \cite{Mawson_2019}.
A particularly striking proposal is anyon superconductivity \cite{Lee_1989,Lee_1991}, in which anyons condense to form a novel superconducting state.

A powerful mean field-theoretic framework for realizing anyon superconductivity is the Chern--Simons--Landau--Ginzburg (CSLG) model.
This theory extends the Ginzburg--Landau (GL) model by the addition of a Chern--Simons (CS) term, which enforces a magnetic flux-charge relation.
The resulting vortices not only carry quantized magnetic flux, as in the standard GL model, but also possess electric charge determined by the CS coupling \cite{Frochlich_1989}.
These charged vortices are anyons, whose statistics are neither bosonic nor fermionic.

The presence of the CS term renders the problem of identifying static energy minimizers intrinsically non-local and mathematically subtle.
In particular, the associated canonical energy functional is not bounded from below, thereby precluding the direct use of standard variational techniques such as gradient descent.
Despite these difficulties, the CSLG model provide a natural setting connecting condensed matter physics, field theory, and mathematics.
On the condensed matter side, they offer an effective description of anyon condensation in the quantum Hall regime \cite{Hansson_2004,Fradkin_1990,Banks_1990,Boyanovsky_1991}.
Mathematically, they are linked to rigorous results on vortex condensation in Chern–Simons–Higgs systems \cite{Caffarelli_1995,Bazeia_2017} and to generalized self-dual models \cite{Andrade_2020,Andrade_2025,Ghosh_1994,Torres_1992,Bazeia_2012,Dunne_1991}.
The interplay between the Yang--Mills and Chern--Simons actions can give rise to rich and unexpected vortex physics.

Challenges of this type are not unique to the CSLG framework, but arise more broadly across both condensed matter and high energy physics.
In condensed matter systems, analogous difficulties occur in the treatment of demagnetization effects in chiral magnets \cite{Leask_Speight_2025} and depolarization fields in chiral liquid crystals \cite{Leask_2025}.
Related complications also emerge in nuclear and particle physics, for example through the Coulomb back-reaction in the nuclear Skyrme model \cite{Gudnason_2025} and in the formation of metastable knotted solitons within extensions of the Standard Model \cite{Hamada_2025}.

In this work, we revisit the CSLG model with the usual quartic Higgs potential.
We first derive the static energy functional and corresponding field equations, emphasizing how Gauss’ law enforces the proportionality between electric charge and magnetic flux, thereby making each vortex an anyon.
We then investigate vortex interactions and demonstrate how the CS level qualitatively alters their long-range behavior.
In particular, we show that even at the Bogomolny point $\lambda=1$, where ordinary Ginzburg--Landau vortices are non-interacting, the presence of a CS term induces repulsion.
More generally, the competition between Higgs attraction and electrostatic repulsion allows for the formation of stable multi-vortex anyon bound states, thereby providing a realization of the hybrid nature of the anyon superconductor.

Traditionally, superconductivity has been classified as type I or type II, with type-1.5 emerging in multicomponent systems that support non-monotonic intervortex forces.
It is known that the typology in quantum Hall liquids (described by the CSLG model without the Yang--Mills term) is richer than that in single-component superconductors, owing to the lack of time-reversal symmetry \cite{Kivelson_2012}.
We demonstrate that the Chern--Simons extension of the Ginzburg--Landau model realizes the same dual-nature in a single-component condensate.
This establishes a new route to hybrid superconductivity, extending the typology beyond the dichotomy of type I/II and multicomponent type-1.5, by showing that topological gauge couplings can generate non-monotonic vortex matter.


\emph{The Ginzburg--Landau model of anyon superconductivity.}
The CSLG model is described by the superconducting order parameter $\psi: \mathbb{R}^{2+1}\rightarrow\mathbb{C}$, also known as the Higgs field, and an abelian gauge field $\vec{A}=(A_0,A_1,A_2)\in\mathbb{R}^{2+1}$.
Associated to the abelian gauge field is the gauge covariant derivative $D_\mu=\partial_\mu + iqA_\mu$, where $q$ is the gauge charge.
The gauge field strength is given by the curvature $F_{\mu\nu}=\partial_\mu A_\nu - \partial_\nu A_\mu$.
From the field strength we define the magnetic field $B=F_{12}$ and the electric field $E_i=F_{0i}$.
We will consider the model defined on Minkowski spacetime $\mathbb{R}^{2+1}$, which is endowed with the Minkowski metric $\eta$ and metric signature $(+--)$.

This GL model of anyon superconductivity is a gauge field theory that exhibits spontaneous symmetry breaking.
The local $U(1)$ invariance is realized by the gauge transformation $A_\mu \mapsto A_\mu +\partial_\mu \alpha(x)$ and $\psi \mapsto\psi e^{i\alpha(x)}$.
The action of the anyonic theory is $S=\int \textup{d}^3x\mathcal{L}$, where the Lagrangian is given by \cite{Hansson_2004,Fradkin_1990,Banks_1990,Boyanovsky_1991}
\begin{equation}
\label{eq: CSLG Lagrangian}
    \mathcal{L} = \frac{1}{2} D_\mu \psi \overline{D^\mu \psi} - \frac{1}{4} F^{\mu\nu} F_{\mu\nu} - V(|\psi|) + \frac{\kappa}{4} \epsilon^{\alpha\beta\gamma} A_\alpha F_{\beta\gamma}.
\end{equation}
The first three terms describe the GL model, also known as the abelian Higgs model in this context, where the second term is the Yang-Mills, or Maxwell, contribution.
The last term is the topological Chern--Simons term,
\begin{equation}
\label{eq: Chern--Simons term}
    \mathcal{L}_{\textup{CS}} = \frac{\kappa}{4} \epsilon^{\alpha\beta\gamma} A_\alpha F_{\beta\gamma} = \frac{\kappa}{2} \left( A_0 B - \epsilon_{ij}A_i E_j \right).
\end{equation}

In this letter we are only interested in static vortex anyons, these are defined as minimizers of the static energy with non-trivial topological degree.
In the GL model, the electric field is absent when considering statics. 
However, due to the presence of the CS term \eqref{eq: Chern--Simons term}, we see that the electric field does not vanish $E_i=F_{0i}=-\partial_i A_0 \neq 0$ and neither does the kinetic term $D_0 \psi \overline{D_0 \psi}=q^2 A_0^2|\psi|^2 \neq 0$.
After an integration by parts, the static Lagrangian can be expressed as
\begin{align}
    \mathcal{L}_{\textup{static}} = \, & \frac{1}{2}(\partial_i A_0)^2 + \kappa A_0 B + \frac{1}{2}q^2 A_0^2|\psi|^2 \nonumber \\
    \, & - \left[ \frac{1}{2} D_i \psi \overline{D_i \psi} + \frac{1}{2}B^2 + V(|\psi|) \right].
\end{align}
Varying the static Lagrangian with respect to the electric potential $A_0$ reveals Gauss' law as an elliptic PDE,
\begin{equation}
\label{eq: Gauss law}
    \left( -\nabla^2 + q^2|\psi|^2 \right) A_0 = -\kappa B.
\end{equation}

Since static vortex anyons are minimizers of the static energy functional, we introduce the static energy of the theory by defining
\begin{align}
\label{eq: Unreduced static energy}
    E = \, & -\int_{\mathbb{R}^2} \textup{d}^2x \, \mathcal{L}_{\textup{static}} \nonumber \\
    = \, & \int_{\mathbb{R}^2} \textup{d}^2x \left\{ \frac{1}{2} D_i \psi \overline{D_i \psi} + \frac{1}{2}B^2 + V(|\psi|) \right. \nonumber \\
    \, & \left. - \frac{1}{2}(\partial_i A_0)^2 - \kappa A_0 B - \frac{1}{2}q^2 A_0^2|\psi|^2 \right\}.
\end{align}
At a first glance this appears not to be bounded from below.
However, the static energy can be transformed into a form that is bounded below and positive (semi-)definite as follows.
Let us take the inner product of Gauss' law \eqref{eq: Gauss law} with the potential $A_0$ and then integrate by parts to obtain
\begin{align}
    -\int_{\mathbb{R}^2} \textup{d}^2x \, \kappa A_0 B = \, & \int_{\mathbb{R}^2} \textup{d}^2x \left[ -A_0 \partial_i \partial_i A_0 + q^2|\psi|^2 A_0^2 \right] \nonumber \\
    = \, & \int_{\mathbb{R}^2} \textup{d}^2x \left[ (\partial_i A_0)^2 + q^2|\psi|^2 A_0^2 \right].
\end{align}
Substituting this into the static energy \eqref{eq: Unreduced static energy} gives an expression for the static energy that is clearly bounded below by 0,
\begin{align}
\label{eq: Positive definite static energy}
    E = \, & \int_{\mathbb{R}^2} \textup{d}^2x \left\{ \frac{1}{2} |\vec{D} \psi|^2 + \frac{1}{2}B^2 + V(|\psi|) + \frac{1}{2}|\vec{\nabla} A_0|^2 \right. \nonumber \\
    \, & \left.  + \frac{1}{2}q^2 A_0^2|\psi|^2 \right\}.
\end{align}
This energy is positive definite and bounded below, therefore it is amenable to gradient descent methods.

To summarize, vortices in this Ginzburg--Landau model of anyon superconductivity are minimizers of the positive definite static energy \eqref{eq: Positive definite static energy}, such that the electromagnetic scalar potential $A_0$ satisfies the non-linear elliptic Gauss constraint \eqref{eq: Gauss law}.


\emph{Anyon superconductivity.}
In (2+1)-dimensions, adding a Chern--Simons term to the action endows particles with fractional statistics.
This means that when you adiabatically braid one particle around another, the many-body wavefunction $\Psi$ picks up a phase
\begin{equation}
    \Psi \rightarrow e^{i(\theta_i+\phi)}\Psi,
\end{equation}
where $\theta_i$ is the intrinsic statistics angle associated with the charge \cite{Lee_1991}.
For bosons $\theta_i=0(\textup{mod}\,2\pi)$ and $\theta_i=\pi(\textup{mod}\,2\pi)$ for fermions.
The other contribution, $\phi$, is a Berry, or Aharonov--Bohm, phase coming from the adiabatic braiding.
Since the Berry phase is added directly to the intrinsic statistics angle $\theta_i$, it fundamentally alters the statistics of the particles.
For example, if we consider fermionic charged particles and $\phi=\pi$, then the resulting statistics of the composite particles is bosonic.
In general, since the Berry phase $\phi$ need not be an integer multiple of $\pi$, it can lead to particles endowed with statistics lying between that of bosons and fermions.
These composite particles are known precisely as anyons.

The braiding phase of the anyons is the Berry phase and is determined by the level (coupling $\kappa$) of the CS term.
For the CS term \eqref{eq: Chern--Simons term}, the Berry phase of the charge-flux composite particles is proportional to
\begin{equation}
    \phi = \frac{\pi q^2}{\kappa}.
\end{equation}
So, we see that the CS level $\kappa$ determines the exchange statistics by setting how much charge binds to flux, and that directly fixes the braiding phase.
This is exactly the same topological mechanism used in the Zhang--Hansson--Kivelson (ZHK) model of the FQHE \cite{ZHK_1989}.
We will now derive this charge-flux relationship explicitly.

In the presence of a CS term the relationship between magnetic flux and electric charge requires special care. 
The starting point is the Gauss constraint \eqref{eq: Gauss law}, which ultimately relates the flux to the charge.
This equation encapsulates the mixing between electric and magnetic fields: a localized magnetic flux distribution acts as a source for the electrostatic potential $A_0$.

The physical electric field is $\vec{E}=-\vec{\nabla}A_0$, so we can compute the electric charge density via the Maxwell equation
\begin{equation}
    \rho_e = \vec{\nabla}\cdot\vec{E} = -\nabla^2 A_0 = -\kappa B - q^2|\psi|^2 A_0,
\end{equation}
where we have used Gauss' law \eqref{eq: Gauss law} to express the charge in terms of the matter and magnetic fields.
Hence, the total physical electric charge $Q_e$ is
\begin{equation}
    Q_e = \int_{\mathbb{R}^2} \textup{d}^2x \, \rho_e = -\kappa \Phi - q^2 \int_{\mathbb{R}^2} \textup{d}^2x\,A_0 |\psi|^2,
\end{equation}
where $\Phi = \int_{\mathbb{R}^2} \textup{d}^2x B$ is the total magnetic flux.
A single magnetic flux quantum is $\Phi_0=2\pi/q$, so the total quantized magnetic flux is $\Phi=N\Phi_0$.
For localized static solutions, $A_0$ decays exponentially and $E_i\to 0$ at spatial infinity.
Therefore, the total electric charge should be zero for localized solutions, $Q_e=0$, which gives the relation
\begin{equation}
    \int_{\mathbb{R}^2} \textup{d}^2x \, q^2A_0 |\psi|^2 = -\kappa\Phi.
\end{equation}

Although the total Maxwell charge $Q_e$ vanishes, the condensate carries a nontrivial internal $U(1)$ charge $Q_m$.
Under a global phase rotation of the scalar, $\psi \mapsto e^{i\alpha}\psi$, the associated Noether current is
\begin{align}
\label{eq: Supercurrent}
    J_\mu = \frac{iq}{2}(\psi \partial_\mu \bar{\psi} - \bar{\psi} \partial_\mu \psi) + q^2 A_\mu |\psi|^2.
\end{align}
From the supercurrent \eqref{eq: Supercurrent}, we obtain the Noether electric charge density
\begin{equation}
    \rho_m=J_0=q^2A_0 |\psi|^2.
\label{eq: Matter charge density}
\end{equation}
Therefore, we find that the magnetic flux $\Phi$ and the Noether electric charge $Q_m$ are related by \cite{Templeton_2000}
\begin{equation}
\label{eq: Noether electric charge}
    Q_m = \int_{\mathbb{R}^2} \textup{d}^2x\,\rho_m = -\kappa \Phi.
\end{equation}
This is the electric charge carried by the condensate relative to the internal $U(1)$ gauge symmetry; it is non-zero and entirely tied to the magnetic flux by Gauss' law \eqref{eq: Gauss law}.
It shows that each unit of magnetic flux carries a quantized amount of Noether charge proportional to the CS level $\kappa$.
This is the hallmark of anyon superconductivity: the non-trivial matter charge \eqref{eq: Matter charge density} signals that vortices in this model are anyonic objects, with each magnetic flux quantum binding a fixed electric charge determined by the CS level.

Importantly, as we have shown, the physical Maxwell charge $Q_e$ vanishes for localized anyons, such that the system remains globally electrically neutral.
In addition to this, both the magnetic flux $\Phi$ and electric charge $Q_m$ are conserved in time.
This follows naturally from the definition of a Noether charge.


\emph{Constrained Newton flow.}
Static vortex anyons are critical points of the static energy \eqref{eq: Unreduced static energy}, so we must solve the associated Euler-Lagrange field equations of the model and also satisfy the Gauss constraint \eqref{eq: Gauss law}.
The Euler-Lagrange field equations are obtained by varying the unreduced static energy functional \eqref{eq: Unreduced static energy} with respect to the Higgs field $\psi$ and the gauge field $(A_1,A_2)$.
This gives us the static Ginzburg--Landau equations
\begin{subequations}
\label{eq: Field equations}
    \begin{align}
        D_i D_i \psi = \, & 2\frac{\partial V}{\partial \bar{\psi}} - q^2A_0^2 \psi, \\
        \partial_j (\partial_j A_i - \partial_i A_j) = \, &  J_i - \kappa \epsilon_{ij}\partial_j A_0,
    \end{align}
\end{subequations}
where the supercurrent is given by \eqref{eq: Supercurrent}.

\begin{figure*}[t]
    \centering
    \includegraphics[width=0.9\textwidth]{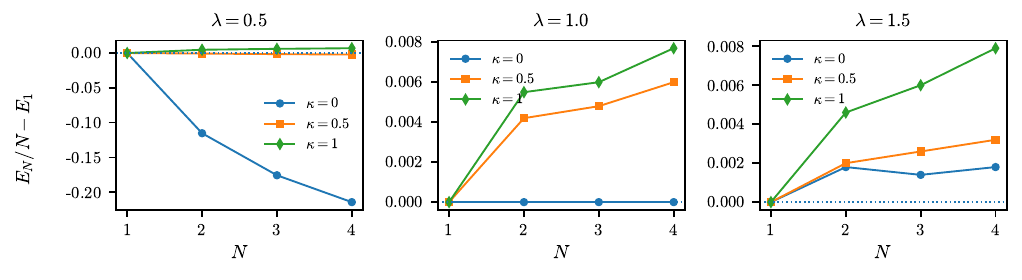}
    \caption{The binding energy per vortex anyon $E_{\textup{bind}}=E_N/N-E_1$ shown for three parameter regimes, corresponding to the value of the Ginzburg--Landau parameter $\lambda$. In the regular Ginzburg--Landau model, the Ginzburg--Landau parameter $\lambda$ controls the type of superconductivity and there is a dichotomy between these types. The first is the type I regime ($\lambda<1$), where the forces between vortices are attractive. Secondly, we have the type II regime ($\lambda>1$), where the intervortex forces are repulsive. At the Bogomolny point ($\lambda=1$), there are no forces between the vortices. This dichotomy is broken in the Chern--Simons extension of the model by introducing an electrostatic repulsion, altering the interaction landscape. Vortex anyons at the Ginzburg--Landau Bogomolny point ($\lambda=1$) now feel a repulsive force. For larger enough Chern--Simons coupling $\kappa$, the type I attractive regime can switch to a repulsive type II regime. Furthermore, a hybrid superconducting state can exist where the vortex anyons experience short-range repulsion and long-range attraction, forming stable bound multi-vortex anyons.}
    \label{fig: Binding energies}
\end{figure*}

Our method for obtaining static vortex anyons is carried out using constrained Newton flow.
This is a method developed originally in the context of high energy physics \cite{Gudnason_2020,Leask_Harland_2024,Leask_2024} and consists of two main features.
The first addresses the Gauss constraint \eqref{eq: Gauss law} by reformulating the constraint as an unconstrained optimization problem.
That is, we want to minimize the functional
\begin{equation}
\label{eq: Unconstrained functional}
    F(A_0) = \int_{\mathbb{R}^2} \textup{d}^2x \left\{ \frac{1}{2}|\vec{\nabla}A_0|^2 + \frac{1}{2} q^2 |\psi|^2 A_0^2 + \kappa B A_0 \right\}
\end{equation}
with respect to the scalar potential $A_0$, while keeping the Higgs field $\psi$ and gauge field $\vec{A}$ fixed.
Clearly, varying \eqref{eq: Unconstrained functional} with respect to $A_0$ yields the Gauss constraint \eqref{eq: Gauss law}.
To solve this unconstrained problem we use non-linear conjugate gradient descent with a line search strategy.
The step-size in the conjugate direction is updated using the Polak--Ribière--Polyak method.

The second part of the algorithm deals with solving the static GL equations \eqref{eq: Field equations}, under the assumption that the scalar potential $A_0$ satisfies the Gauss constraint \eqref{eq: Gauss law}.
This is achieved using an accelerated gradient descent method with flow arresting criteria, known as arrested Newton flow.
We formulate the minimization as a second order dynamical problem and solve the second order system
\begin{align}
    \frac{\textup{d}^2\psi}{\textup{d}t^2} = \, & \frac{1}{2}D_i D_i \psi - \frac{\partial V}{\partial \bar{\psi}} + \frac{1}{2}q^2A_0^2 \psi, \\
    \frac{\textup{d}^2 A_i}{\textup{d}t^2} = \, & \partial_j (\partial_j A_i - \partial_i A_j) - J_i + \kappa \epsilon_{ij}\partial_j A_0,
\end{align}
where $t$ is a fictitious time coordinate and with some appropriate initial configuration $\{\psi(0),\vec{A}(0)\}=\{\psi_0,\vec{A}_0\}$.
This system can then be reduced to a coupled first order system, which we solve using a fourth order Runge-Kutta method.

The constrained Newton flow algorithm proceeds by evolving the Higgs and gauge fields using the arrested Newton flow method, starting from some initial configuration.
During every iteration of the Newton flow method, we ensure the the Gauss constraint \eqref{eq: Gauss law} is satisfied by using the non-linear conjugate gradient method detailed above.
In addition, we also check to see if the total energy of the system has increased over the time step $\delta t$.
If $E(t+\delta t) > E(t)$, we arrest the flow by taking out all of the kinetic energy in the system and restart the flow from $t+\delta t$.
This algorithm naturally converges to a local minimum must faster than regular gradient descent.

To implement the constrained Newton flow algorithm, we employ a fourth order central finite difference method.
This is performed on a meshgrid of $N^2$ grid points with grid spacing $h$ and results in a discrete approximation $E_{\textup{dis}}$ to the static energy $E$.
The flow is deemed to have converged when $\lVert E_{\textup{dis}} \rVert_\infty<\epsilon$, where $\epsilon$ is some threshold tolerance (we use $\epsilon=10^{-6}$).
We perform this using custom high-performance computing software, developed for GPU architecture.

In order to ensure the algorithm converges to a multi-vortex configuration, we must use an appropriate initial configuration.
We choose to use the axially symmetric vortex ansatz \cite{Nielsen_1973,Hindmarsh_1995}
\begin{equation}
\label{eq: Nielsen-Olesen ansatz}
    \begin{split}
        \psi =  m\phi(r)e^{iN\theta}, \quad\vec{A} =  \frac{Na(r)}{qr}\left(\sin\theta,-\cos\theta\right),
    \end{split}
\end{equation}
where $N$ is the number of quasi-particles (vortex number) and $m$ the ground state value of the Higgs field.
The profile functions are monotonically increasing functions that satisfy the boundary conditions $\phi(0)=a(0)=0$ and 
\begin{equation}
    \lim_{r\rightarrow\infty}\phi(r)=\lim_{r\rightarrow\infty}a(r)=1.
\end{equation}
For our initial configuration, we simply set $\phi_0(r)=a_0(r)=\tanh(r)$ and $A_0=0$.
The magnetic flux of such an $N$-vortex configuration is computed to be given by
\begin{equation}
    \Phi = \int_{\mathbb{R}^2} \textup{d}^2x B = \frac{2\pi N}{q}=N\Phi_0.
\end{equation}


\emph{Multi-vortex anyon bound states.}
The inclusion of the CS term has a profound effect on the binding properties of vortices.
Consider the conventional GL model with quartic potential
\begin{equation}
\label{eq: Higgs potential}
    V(|\psi|) = \frac{\lambda}{8} \left( m^2-|\psi|^2 \right)^2,
\end{equation}
where the Higgs mass is $m_H=\sqrt{\lambda}m$ and $\lambda$ is the GL parameter dictating the type of superconductivity.
The interaction between vortices in the GL model is controlled by the relative size of the Higgs coherence length $\xi_H=1/m_H$ and the magnetic penetration depth $\lambda_H=1/m_A$, with gauge mass $m_A=qm$.
For $\lambda<1$ the Higgs mode is lighter than the gauge mode and vortices attract (type I).
For $\lambda>1$, vortices repel (type II).
At the Bogomolny point $\lambda=1$, the two modes are degenerate and the inter-vortex forces cancel.

In the CSLG model, this dichotomy is altered.
Each vortex carries not only magnetic flux but also a Noether electric charge tied to the flux by Gauss’ law \eqref{eq: Gauss law}.
The resulting electrostatic interaction contributes an additional repulsive channel.
Numerical minimization of the static energy \eqref{eq: Positive definite static energy} with constrained Newton flow reveals that for $\kappa >0$ the interaction behavior of vortices can change qualitatively.
For small CS coupling $\kappa$, vortices in the $\lambda<1$ regime still experience overall attraction and tend to coalesce, resembling type I behavior.
At intermediate $\kappa$, the long-range Higgs-mediated attraction remains dominant, but the electrostatic repulsion becomes significant at short distances.
This frustrates collapse into a single axially symmetric multi-vortex and instead favors the formation of spatially separated bound states with negative binding energy.
At larger $\kappa$, the repulsion dominates at all scales, even in the nominal type I regime, driving vortices apart and mimicking type II superconductivity.
The resulting binding energies of multi-vortex anyons in the three GL parameter $\lambda$ regimes, for various CS couplings $\kappa$, is shown in Fig. ~\ref{fig: Binding energies}.

\begin{figure*}[t]
    \centering
    \includegraphics[width=0.8\textwidth]{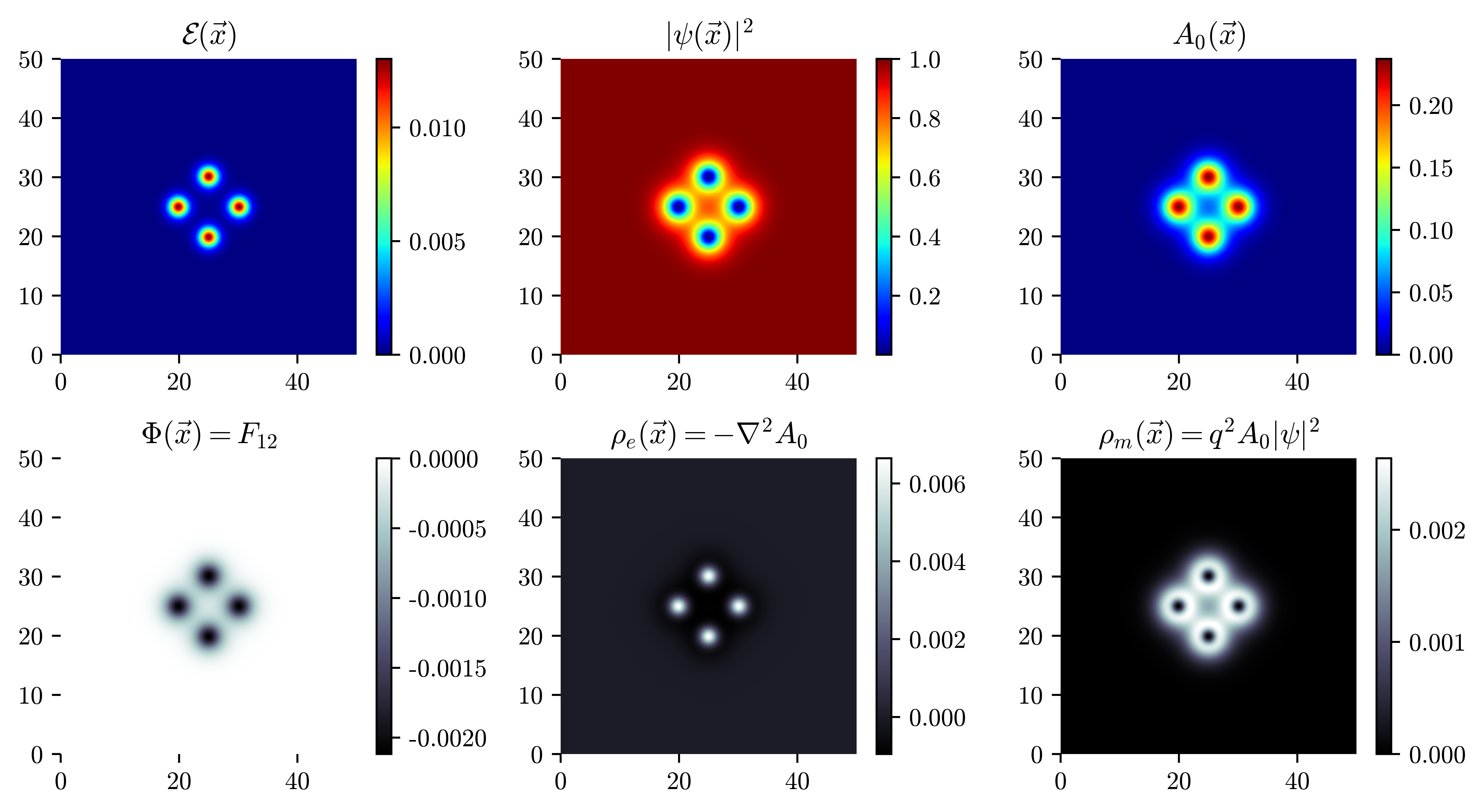}
    \caption{Density plots of an $N=4$ vortex anyon exhibiting hybrid anyon superconductivity. Each individual vortex carries a magnetic flux $\Phi_0=-2\pi$ and an electric charge $-\kappa\Phi_0$, with $\kappa=\tfrac12$. It can be seen that the magnetic flux is centered on the cores of the vortices (bottom left panel), and the electric Noether charge forms a ring of charge around each vortex core (bottom right panel). These vortex anyons are obtained for the Higgs potential \eqref{eq: Higgs potential}, with parameters $m=1$ and $\lambda=\tfrac12$. The binding energy of this $4$-vortex is negative and, so, the vortices form a bound state. They do not collapse to form an axially symmetric state with coincident cores as they experience short-range repulsion.}
    \label{fig: Type 1.5}
\end{figure*}

The intermediate hybrid regime, where vortex anyons repel at short range but attract at longer range, is similar to that of type 1.5 superconductivity \cite{Babaev_2002,Babaev_Speight_2005,Babaev_Carlstrom_2010}.
This phenomenon is illustrated in Fig. \ref{fig: Type 1.5}, where an $N=4$ configuration relaxes into a stable non-axially symmetric bound state: vortex anyons remain separated due to short-range repulsion yet are held together by the residual long-range attraction.
Each constituent vortex carries magnetic flux $\Phi_0=2\pi/q$ and a compensating Noether charge $-\kappa\Phi_0$, and the multi-vortex anyon cluster retains overall electric charge neutrality in the Maxwell sense.

The stability of such bound states has important implications.
Unlike type I or II, where vortices either collapse into a single giant core or form an Abrikosov lattice, anyon systems can support molecular-like vortex clusters.
These bound states could underpin new collective phases of anyon superconductors, distinct from both the homogeneous condensates of conventional superconductivity and from the crystalline order of type II.
Their existence also strengthens the analogy with multi-component superconductors, where competing length scales naturally give rise to hybrid physics \cite{Babaev_2011,Babaev_2017}.

\emph{Hybrid superconductivity.}
To explain the observation of the hybrid superconductivity in our single-component model, we consider the asymptotic tails of the vortex anyons and their long-range interactions.
That is, we expand about the ground state configuration $\{\psi,A_\mu\}=\{m+\phi,0+a_\mu\}$ and consider field equations linear in the perturbations $\{\phi,a_\mu\}$.
We find that the magnetic field and electric field must satisfy the same linearized field equations, $\Delta_\kappa B = 0$ and $\Delta_\kappa E_i = 0$, where
\begin{equation}
    \Delta_\kappa=\left[ \left( -\nabla^2 + m_A^2 \right)^2 + \kappa^2\nabla^2 \right].
\end{equation}
This operator can be factorized into complex-conjugate eigenmodes $m_\pm$, which are the static screening masses, given by
\begin{equation}
    m_\pm = \sqrt{m_A^2 - \frac{\kappa^2}{4}} \pm i\frac{\kappa}{2}.
\end{equation}
Therefore, we see that the magnetic and electric fields share a common penetration depth $\lambda_{\textup{gauge}}$ but differ by an oscillation frequency $1/\lambda_{\textup{osc}}$, with
\begin{equation}
    \lambda_{\textup{gauge}} = \frac{1}{\sqrt{m_A^2-\kappa^2/4}}, \quad \lambda_{\textup{osc}} = \frac{4\pi}{\kappa}.
\end{equation}
That is, asymptotically, the magnetic and electric fields have a damped oscillatory profile and share the same decay rate.
The real dynamical gauge masses of the theory are obtained from the static screening masses by a Wick rotation $\kappa\mapsto i\kappa$.

The long-range oscillation behavior of the gauge modes leads to more interesting long-range interactions.
We find that the interaction energy of a pair of separated vortex anyons is determined by
\begin{align}
    V_{\textup{int}}(R) = \, & 2\pi |c_B|^2 \sqrt{\frac{2\pi}{m_A R}} \, e^{-\sqrt{m_A^2 - \frac{\kappa^2}{4}} R} \cos\left( \frac{\kappa}{2} R - \gamma \right)  \nonumber \\
    \, & - 2\pi c_H^2 K_0(m_H R).
\end{align}
The standard scalar superconducting order parameter contribution remains monotone attractive, as in the regular GL model.
However, the gauge term becomes a damped oscillator with envelope $e^{-\alpha R}/\sqrt{R}$, decay rate $\alpha=\tfrac12\sqrt{4m_A^2-\kappa^2}$, and oscillation frequency $\beta=\kappa/2$, alternating between attractive and repulsive behavior.
If the gauge term dominates at a shorter range than the Higgs term, it can provide a repulsive force initially ($R<\tfrac{\pi}{\kappa}+\tfrac{2\gamma}{\kappa}$) before switching to an attractive force at longer range ($R>\tfrac{\pi}{\kappa}+\tfrac{2\gamma}{\kappa}$), and repeating this behavior in a decaying oscillatory fashion as the separation distance $R$ increases.
In contrast to type 1.5 superconductivity in multiband superconductors, where the hybrid behavior arises to due competing length scales with $\xi_1 < \lambda<\xi_2$, this long-range oscillatory behavior can give rise to the shorter-range repulsion and long-range attraction essential for hybrid superconductivity.


\emph{Conclusion.}
In this work we have examined vortex physics in the Chern--Simons--Landau--Ginzburg model, highlighting how the addition of a CS term profoundly modifies the landscape of superconductivity in two spatial dimensions.
Gauss’ law ties each magnetic flux quantum to a compensating Noether electric charge, rendering vortices intrinsically anyonic.
This charge-flux attachment mechanism introduces an electrostatic contribution to the vortex interaction and leads to non-trivial screening behavior.
Unlike the Ginzburg--Landau model, where vortices exhibit a simple type I/II dichotomy determined by the GL parameter, the CSLG model yields complex-conjugate gauge screening masses.
Consequently, the electric and magnetic fields decay with a common penetration depth but acquire an oscillatory phase shift.

Our numerical simulations, carried out via the constrained Newton flow algorithm, demonstrate that this altered screening structure allows for stable multi-vortex anyon bound states.
These bound states exhibit short-range repulsion and long-range attraction, the defining characteristic of hybrid superconductivity.
We have thereby provided a theoretical realization of hybrid behavior in an anyon superconductor.

We note that the dual superconducting nature of the anyon superconductor is not unique to this single component model, other single component superconducting systems can give rise to non-monotonic intervortex forces.
For example, in non-centrosymmetric superconductors, broken inversion symmetry produces spiral magnetic field decay \cite{Samoilenka_2020} and parity-breaking terms can cause magnetic field inversion \cite{Garaud_2020}.
These both lead to non-monotonic intervortex forces that stabilize vortex bound states.
This is in close analogy to how the Chern--Simons term in our model generates oscillatory gauge tails and dual-nature vortex interactions.

The emergence of molecular-like vortex anyon clusters suggests that anyon superconductivity may host a much richer set of collective states than previously appreciated.
Recent works have also explored related anyonic and topological mechanisms in two-dimensional systems: anyonic molecules in fractional quantum Hall liquids \cite{Heras_2020}, doped fractional quantum anomalous Hall insulators exhibiting itinerant anyon superconductivity \cite{Shi_2025}, and chiral pseudospin liquids in moiré heterostructures \cite{Kuhlenkamp_2024}.
These studies underscore the growing interest in anyon-mediated condensation and topological gauge responses, providing complementary microscopic contexts for the effective Chern--Simons mechanism and hybrid vortex behavior examined here.
In addition to extending the classification of superconducting phases, these results highlight deep connections between topological gauge theories, fractional statistics, and non-standard superconductivity.
Future directions include a systematic study of vortex lattice phases in this model \cite{Speight_Winyard_2025}, as well as the role of the CS term in dynamical interactions of vortex anyons \cite{Bazeia_2024}, and the short-range interactions of vortex anyons \cite{Winyard_2025}.


\emph{Acknowledgments.}
I would like to thank E. Babaev, A. Talkachov and M. Speight for useful discussions.
The author acknowledges funding from the Olle Engkvists Stiftelse through the grant 226-0103.


\bibliography{main.bib}

\begin{thebibliography}{51}%
\makeatletter
\providecommand \@ifxundefined [1]{%
 \@ifx{#1\undefined}
}%
\providecommand \@ifnum [1]{%
 \ifnum #1\expandafter \@firstoftwo
 \else \expandafter \@secondoftwo
 \fi
}%
\providecommand \@ifx [1]{%
 \ifx #1\expandafter \@firstoftwo
 \else \expandafter \@secondoftwo
 \fi
}%
\providecommand \natexlab [1]{#1}%
\providecommand \enquote  [1]{``#1''}%
\providecommand \bibnamefont  [1]{#1}%
\providecommand \bibfnamefont [1]{#1}%
\providecommand \citenamefont [1]{#1}%
\providecommand \href@noop [0]{\@secondoftwo}%
\providecommand \href [0]{\begingroup \@sanitize@url \@href}%
\providecommand \@href[1]{\@@startlink{#1}\@@href}%
\providecommand \@@href[1]{\endgroup#1\@@endlink}%
\providecommand \@sanitize@url [0]{\catcode `\\12\catcode `\$12\catcode `\&12\catcode `\#12\catcode `\^12\catcode `\_12\catcode `\%12\relax}%
\providecommand \@@startlink[1]{}%
\providecommand \@@endlink[0]{}%
\providecommand \url  [0]{\begingroup\@sanitize@url \@url }%
\providecommand \@url [1]{\endgroup\@href {#1}{\urlprefix }}%
\providecommand \urlprefix  [0]{URL }%
\providecommand \Eprint [0]{\href }%
\providecommand \doibase [0]{https://doi.org/}%
\providecommand \selectlanguage [0]{\@gobble}%
\providecommand \bibinfo  [0]{\@secondoftwo}%
\providecommand \bibfield  [0]{\@secondoftwo}%
\providecommand \translation [1]{[#1]}%
\providecommand \BibitemOpen [0]{}%
\providecommand \bibitemStop [0]{}%
\providecommand \bibitemNoStop [0]{.\EOS\space}%
\providecommand \EOS [0]{\spacefactor3000\relax}%
\providecommand \BibitemShut  [1]{\csname bibitem#1\endcsname}%
\let\auto@bib@innerbib\@empty
\bibitem [{\citenamefont {Zhang}\ \emph {et~al.}(1989)\citenamefont {Zhang}, \citenamefont {Hansson},\ and\ \citenamefont {Kivelson}}]{ZHK_1989}%
  \BibitemOpen
  \bibfield  {author} {\bibinfo {author} {\bibfnamefont {S.~C.}\ \bibnamefont {Zhang}}, \bibinfo {author} {\bibfnamefont {T.~H.}\ \bibnamefont {Hansson}},\ and\ \bibinfo {author} {\bibfnamefont {S.}~\bibnamefont {Kivelson}},\ }\bibfield  {title} {\bibinfo {title} {Effective-field-theory model for the fractional quantum {Hall} effect},\ }\href {https://doi.org/10.1103/PhysRevLett.62.82} {\bibfield  {journal} {\bibinfo  {journal} {Phys. Rev. Lett.}\ }\textbf {\bibinfo {volume} {62}},\ \bibinfo {pages} {82} (\bibinfo {year} {1989})}\BibitemShut {NoStop}%
\bibitem [{\citenamefont {Zhang}(1992)}]{Zhang_1992}%
  \BibitemOpen
  \bibfield  {author} {\bibinfo {author} {\bibfnamefont {S.~C.}\ \bibnamefont {Zhang}},\ }\bibfield  {title} {\bibinfo {title} {{The Chern-Simons-Landau-Ginzburg theory of the fractional quantum Hall effect}},\ }\href {https://doi.org/10.1142/S0217979292000037} {\bibfield  {journal} {\bibinfo  {journal} {Int. J. Mod. Phys. B}\ }\textbf {\bibinfo {volume} {6}},\ \bibinfo {pages} {25} (\bibinfo {year} {1992})}\BibitemShut {NoStop}%
\bibitem [{\citenamefont {Mulligan}\ \emph {et~al.}(2010)\citenamefont {Mulligan}, \citenamefont {Nayak},\ and\ \citenamefont {Kachru}}]{Mulligan_2010}%
  \BibitemOpen
  \bibfield  {author} {\bibinfo {author} {\bibfnamefont {M.}~\bibnamefont {Mulligan}}, \bibinfo {author} {\bibfnamefont {C.}~\bibnamefont {Nayak}},\ and\ \bibinfo {author} {\bibfnamefont {S.}~\bibnamefont {Kachru}},\ }\bibfield  {title} {\bibinfo {title} {Isotropic to anisotropic transition in a fractional quantum {Hall} state},\ }\href {https://doi.org/10.1103/PhysRevB.82.085102} {\bibfield  {journal} {\bibinfo  {journal} {Phys. Rev. B}\ }\textbf {\bibinfo {volume} {82}},\ \bibinfo {pages} {085102} (\bibinfo {year} {2010})}\BibitemShut {NoStop}%
\bibitem [{\citenamefont {Mawson}\ \emph {et~al.}(2019)\citenamefont {Mawson}, \citenamefont {Petersen}, \citenamefont {Slingerland},\ and\ \citenamefont {Simula}}]{Mawson_2019}%
  \BibitemOpen
  \bibfield  {author} {\bibinfo {author} {\bibfnamefont {T.}~\bibnamefont {Mawson}}, \bibinfo {author} {\bibfnamefont {T.~C.}\ \bibnamefont {Petersen}}, \bibinfo {author} {\bibfnamefont {J.~K.}\ \bibnamefont {Slingerland}},\ and\ \bibinfo {author} {\bibfnamefont {T.~P.}\ \bibnamefont {Simula}},\ }\bibfield  {title} {\bibinfo {title} {Braiding and fusion of non-abelian vortex anyons},\ }\href {https://doi.org/10.1103/PhysRevLett.123.140404} {\bibfield  {journal} {\bibinfo  {journal} {Phys. Rev. Lett.}\ }\textbf {\bibinfo {volume} {123}},\ \bibinfo {pages} {140404} (\bibinfo {year} {2019})}\BibitemShut {NoStop}%
\bibitem [{\citenamefont {Lee}\ and\ \citenamefont {Fisher}(1989)}]{Lee_1989}%
  \BibitemOpen
  \bibfield  {author} {\bibinfo {author} {\bibfnamefont {D.-H.}\ \bibnamefont {Lee}}\ and\ \bibinfo {author} {\bibfnamefont {M.~P.~A.}\ \bibnamefont {Fisher}},\ }\bibfield  {title} {\bibinfo {title} {Anyon superconductivity and the fractional quantum {Hall} effect},\ }\href {https://doi.org/10.1103/PhysRevLett.63.903} {\bibfield  {journal} {\bibinfo  {journal} {Phys. Rev. Lett.}\ }\textbf {\bibinfo {volume} {63}},\ \bibinfo {pages} {903} (\bibinfo {year} {1989})}\BibitemShut {NoStop}%
\bibitem [{\citenamefont {Lee}\ and\ \citenamefont {Fisher}(1991)}]{Lee_1991}%
  \BibitemOpen
  \bibfield  {author} {\bibinfo {author} {\bibfnamefont {D.-H.}\ \bibnamefont {Lee}}\ and\ \bibinfo {author} {\bibfnamefont {M.~P.~A.}\ \bibnamefont {Fisher}},\ }\bibfield  {title} {\bibinfo {title} {Anyon superconductivity and charge-vortex duality},\ }\href {https://doi.org/10.1142/S0217979291001061} {\bibfield  {journal} {\bibinfo  {journal} {Int. J. Mod. Phys. B}\ }\textbf {\bibinfo {volume} {05}},\ \bibinfo {pages} {2675} (\bibinfo {year} {1991})}\BibitemShut {NoStop}%
\bibitem [{\citenamefont {Fr\"ohlich}\ and\ \citenamefont {Marchetti}(1989)}]{Frochlich_1989}%
  \BibitemOpen
  \bibfield  {author} {\bibinfo {author} {\bibfnamefont {J.}~\bibnamefont {Fr\"ohlich}}\ and\ \bibinfo {author} {\bibfnamefont {P.}~\bibnamefont {Marchetti}},\ }\bibfield  {title} {\bibinfo {title} {Quantum field theories of vortices and anyons},\ }\href {https://doi.org/10.1007/BF01217803} {\bibfield  {journal} {\bibinfo  {journal} {Commun. Math. Phys.}\ }\textbf {\bibinfo {volume} {121}},\ \bibinfo {pages} {177} (\bibinfo {year} {1989})}\BibitemShut {NoStop}%
\bibitem [{\citenamefont {Hansson}\ \emph {et~al.}(2004)\citenamefont {Hansson}, \citenamefont {Oganesyan},\ and\ \citenamefont {Sondhi}}]{Hansson_2004}%
  \BibitemOpen
  \bibfield  {author} {\bibinfo {author} {\bibfnamefont {T.}~\bibnamefont {Hansson}}, \bibinfo {author} {\bibfnamefont {V.}~\bibnamefont {Oganesyan}},\ and\ \bibinfo {author} {\bibfnamefont {S.}~\bibnamefont {Sondhi}},\ }\bibfield  {title} {\bibinfo {title} {Superconductors are topologically ordered},\ }\href {https://doi.org/10.1016/j.aop.2004.05.006} {\bibfield  {journal} {\bibinfo  {journal} {Ann. Phys.}\ }\textbf {\bibinfo {volume} {313}},\ \bibinfo {pages} {497} (\bibinfo {year} {2004})}\BibitemShut {NoStop}%
\bibitem [{\citenamefont {Fradkin}(1990)}]{Fradkin_1990}%
  \BibitemOpen
  \bibfield  {author} {\bibinfo {author} {\bibfnamefont {E.}~\bibnamefont {Fradkin}},\ }\bibfield  {title} {\bibinfo {title} {Superfluidity of the lattice anyon gas and topological invariance},\ }\href {https://doi.org/10.1103/PhysRevB.42.570} {\bibfield  {journal} {\bibinfo  {journal} {Phys. Rev. B}\ }\textbf {\bibinfo {volume} {42}},\ \bibinfo {pages} {570} (\bibinfo {year} {1990})}\BibitemShut {NoStop}%
\bibitem [{\citenamefont {Banks}\ and\ \citenamefont {Lykken}(1990)}]{Banks_1990}%
  \BibitemOpen
  \bibfield  {author} {\bibinfo {author} {\bibfnamefont {T.}~\bibnamefont {Banks}}\ and\ \bibinfo {author} {\bibfnamefont {J.~D.}\ \bibnamefont {Lykken}},\ }\bibfield  {title} {\bibinfo {title} {{Landau-Ginzburg} description of anyonic superconductors},\ }\href {https://doi.org/10.1016/0550-3213(90)90439-K} {\bibfield  {journal} {\bibinfo  {journal} {Nucl. Phys. B}\ }\textbf {\bibinfo {volume} {336}},\ \bibinfo {pages} {500} (\bibinfo {year} {1990})}\BibitemShut {NoStop}%
\bibitem [{\citenamefont {Boyanovsky}(1991)}]{Boyanovsky_1991}%
  \BibitemOpen
  \bibfield  {author} {\bibinfo {author} {\bibfnamefont {D.}~\bibnamefont {Boyanovsky}},\ }\bibfield  {title} {\bibinfo {title} {Vortices in {Landau-Ginzburg} theories of anyonic superconductivity},\ }\href {https://doi.org/10.1016/0550-3213(91)90168-W} {\bibfield  {journal} {\bibinfo  {journal} {Nucl. Phys. B}\ }\textbf {\bibinfo {volume} {350}},\ \bibinfo {pages} {906} (\bibinfo {year} {1991})}\BibitemShut {NoStop}%
\bibitem [{\citenamefont {Caffarelli}\ and\ \citenamefont {Yang}(1995)}]{Caffarelli_1995}%
  \BibitemOpen
  \bibfield  {author} {\bibinfo {author} {\bibfnamefont {L.~A.}\ \bibnamefont {Caffarelli}}\ and\ \bibinfo {author} {\bibfnamefont {Y.}~\bibnamefont {Yang}},\ }\bibfield  {title} {\bibinfo {title} {{Vortex condensation in the Chern--Simons Higgs model: an existence theorem}},\ }\href {https://doi.org/10.1007/BF02101552} {\bibfield  {journal} {\bibinfo  {journal} {Commun. Math. Phys.}\ }\textbf {\bibinfo {volume} {168}},\ \bibinfo {pages} {321} (\bibinfo {year} {1995})}\BibitemShut {NoStop}%
\bibitem [{\citenamefont {Bazeia}\ \emph {et~al.}(2017)\citenamefont {Bazeia}, \citenamefont {Losano}, \citenamefont {Marques},\ and\ \citenamefont {Menezes}}]{Bazeia_2017}%
  \BibitemOpen
  \bibfield  {author} {\bibinfo {author} {\bibfnamefont {D.}~\bibnamefont {Bazeia}}, \bibinfo {author} {\bibfnamefont {L.}~\bibnamefont {Losano}}, \bibinfo {author} {\bibfnamefont {M.}~\bibnamefont {Marques}},\ and\ \bibinfo {author} {\bibfnamefont {R.}~\bibnamefont {Menezes}},\ }\bibfield  {title} {\bibinfo {title} {{Compact Chern–Simons vortices}},\ }\href {https://doi.org/10.1016/j.physletb.2017.06.055} {\bibfield  {journal} {\bibinfo  {journal} {Phys. Lett. B}\ }\textbf {\bibinfo {volume} {772}},\ \bibinfo {pages} {253} (\bibinfo {year} {2017})}\BibitemShut {NoStop}%
\bibitem [{\citenamefont {Andrade}\ \emph {et~al.}(2020)\citenamefont {Andrade}, \citenamefont {Bazeia}, \citenamefont {Marques},\ and\ \citenamefont {Menezes}}]{Andrade_2020}%
  \BibitemOpen
  \bibfield  {author} {\bibinfo {author} {\bibfnamefont {I.}~\bibnamefont {Andrade}}, \bibinfo {author} {\bibfnamefont {D.}~\bibnamefont {Bazeia}}, \bibinfo {author} {\bibfnamefont {M.~A.}\ \bibnamefont {Marques}},\ and\ \bibinfo {author} {\bibfnamefont {R.}~\bibnamefont {Menezes}},\ }\bibfield  {title} {\bibinfo {title} {Vortices in {Maxwell-Chern-Simons-Higgs} models with nonminimal coupling},\ }\href {https://doi.org/10.1103/PhysRevD.102.045018} {\bibfield  {journal} {\bibinfo  {journal} {Phys. Rev. D}\ }\textbf {\bibinfo {volume} {102}},\ \bibinfo {pages} {045018} (\bibinfo {year} {2020})}\BibitemShut {NoStop}%
\bibitem [{\citenamefont {Andrade}\ \emph {et~al.}(2025)\citenamefont {Andrade}, \citenamefont {Casana},\ and\ \citenamefont {da~Hora}}]{Andrade_2025}%
  \BibitemOpen
  \bibfield  {author} {\bibinfo {author} {\bibfnamefont {J.}~\bibnamefont {Andrade}}, \bibinfo {author} {\bibfnamefont {R.}~\bibnamefont {Casana}},\ and\ \bibinfo {author} {\bibfnamefont {E.}~\bibnamefont {da~Hora}},\ }\bibfield  {title} {\bibinfo {title} {{BPS chiral vortices in Maxwell-Higgs electrodynamics}},\ }\href {https://doi.org/10.1103/PhysRevD.111.036019} {\bibfield  {journal} {\bibinfo  {journal} {Phys. Rev. D}\ }\textbf {\bibinfo {volume} {111}},\ \bibinfo {pages} {036019} (\bibinfo {year} {2025})}\BibitemShut {NoStop}%
\bibitem [{\citenamefont {Ghosh}(1994)}]{Ghosh_1994}%
  \BibitemOpen
  \bibfield  {author} {\bibinfo {author} {\bibfnamefont {P.~K.}\ \bibnamefont {Ghosh}},\ }\bibfield  {title} {\bibinfo {title} {Bogomol'nyi equations of {Maxwell-Chern-Simons} vortices from a generalized abelian {Higgs} model},\ }\href {https://doi.org/10.1103/PhysRevD.49.5458} {\bibfield  {journal} {\bibinfo  {journal} {Phys. Rev. D}\ }\textbf {\bibinfo {volume} {49}},\ \bibinfo {pages} {5458} (\bibinfo {year} {1994})}\BibitemShut {NoStop}%
\bibitem [{\citenamefont {Torres}(1992)}]{Torres_1992}%
  \BibitemOpen
  \bibfield  {author} {\bibinfo {author} {\bibfnamefont {M.}~\bibnamefont {Torres}},\ }\bibfield  {title} {\bibinfo {title} {Bogomol'nyi limit for nontopological solitons in a {Chern-Simons} model with anomalous magnetic moment},\ }\href {https://doi.org/10.1103/PhysRevD.46.R2295} {\bibfield  {journal} {\bibinfo  {journal} {Phys. Rev. D}\ }\textbf {\bibinfo {volume} {46}},\ \bibinfo {pages} {R2295} (\bibinfo {year} {1992})}\BibitemShut {NoStop}%
\bibitem [{\citenamefont {Bazeia}\ \emph {et~al.}(2012)\citenamefont {Bazeia}, \citenamefont {Casana}, \citenamefont {da~Hora},\ and\ \citenamefont {Menezes}}]{Bazeia_2012}%
  \BibitemOpen
  \bibfield  {author} {\bibinfo {author} {\bibfnamefont {D.}~\bibnamefont {Bazeia}}, \bibinfo {author} {\bibfnamefont {R.}~\bibnamefont {Casana}}, \bibinfo {author} {\bibfnamefont {E.}~\bibnamefont {da~Hora}},\ and\ \bibinfo {author} {\bibfnamefont {R.}~\bibnamefont {Menezes}},\ }\bibfield  {title} {\bibinfo {title} {Generalized self-dual {Maxwell-Chern-Simons-Higgs} model},\ }\href {https://doi.org/10.1103/PhysRevD.85.125028} {\bibfield  {journal} {\bibinfo  {journal} {Phys. Rev. D}\ }\textbf {\bibinfo {volume} {85}},\ \bibinfo {pages} {125028} (\bibinfo {year} {2012})}\BibitemShut {NoStop}%
\bibitem [{\citenamefont {Dunne}\ and\ \citenamefont {Trugenberger}(1991)}]{Dunne_1991}%
  \BibitemOpen
  \bibfield  {author} {\bibinfo {author} {\bibfnamefont {G.~V.}\ \bibnamefont {Dunne}}\ and\ \bibinfo {author} {\bibfnamefont {C.~A.}\ \bibnamefont {Trugenberger}},\ }\bibfield  {title} {\bibinfo {title} {Self-duality and nonrelativistic {Maxwell-Chern-Simons} solitons},\ }\href {https://doi.org/10.1103/PhysRevD.43.1323} {\bibfield  {journal} {\bibinfo  {journal} {Phys. Rev. D}\ }\textbf {\bibinfo {volume} {43}},\ \bibinfo {pages} {1323} (\bibinfo {year} {1991})}\BibitemShut {NoStop}%
\bibitem [{\citenamefont {Leask}\ and\ \citenamefont {Speight}(2025)}]{Leask_Speight_2025}%
  \BibitemOpen
  \bibfield  {author} {\bibinfo {author} {\bibfnamefont {P.}~\bibnamefont {Leask}}\ and\ \bibinfo {author} {\bibfnamefont {M.}~\bibnamefont {Speight}},\ }\href {https://arxiv.org/abs/2504.17772} {\bibinfo {title} {Demagnetization in micromagnetics: magnetostatic self-interactions of bulk chiral magnetic skyrmions}} (\bibinfo {year} {2025}),\ \Eprint {https://arxiv.org/abs/2504.17772} {arXiv:2504.17772 [cond-mat.mes-hall]} \BibitemShut {NoStop}%
\bibitem [{\citenamefont {Leask}(2025)}]{Leask_2025}%
  \BibitemOpen
  \bibfield  {author} {\bibinfo {author} {\bibfnamefont {P.}~\bibnamefont {Leask}},\ }\bibfield  {title} {\bibinfo {title} {Topological transition from a hopfion to a toron via flexoelectric self-polarization in chiral liquid crystals},\ }\href {https://doi.org/10.1103/gy6m-m7ck} {\bibfield  {journal} {\bibinfo  {journal} {Phys. Rev. Res.}\ }\textbf {\bibinfo {volume} {7}},\ \bibinfo {pages} {043001} (\bibinfo {year} {2025})}\BibitemShut {NoStop}%
\bibitem [{\citenamefont {Gudnason}\ and\ \citenamefont {Speight}(2025)}]{Gudnason_2025}%
  \BibitemOpen
  \bibfield  {author} {\bibinfo {author} {\bibfnamefont {S.~B.}\ \bibnamefont {Gudnason}}\ and\ \bibinfo {author} {\bibfnamefont {J.~M.}\ \bibnamefont {Speight}},\ }\bibfield  {title} {\bibinfo {title} {{Backreacted Coulomb energy in the Skyrme model}},\ }\href {https://doi.org/10.1007/JHEP01(2025)150} {\bibfield  {journal} {\bibinfo  {journal} {J. High Energ. Phys.}\ }\textbf {\bibinfo {volume} {01}},\ \bibinfo {pages} {150}}\BibitemShut {NoStop}%
\bibitem [{\citenamefont {Eto}\ \emph {et~al.}(2025)\citenamefont {Eto}, \citenamefont {Hamada},\ and\ \citenamefont {Nitta}}]{Hamada_2025}%
  \BibitemOpen
  \bibfield  {author} {\bibinfo {author} {\bibfnamefont {M.}~\bibnamefont {Eto}}, \bibinfo {author} {\bibfnamefont {Y.}~\bibnamefont {Hamada}},\ and\ \bibinfo {author} {\bibfnamefont {M.}~\bibnamefont {Nitta}},\ }\bibfield  {title} {\bibinfo {title} {Tying knots in particle physics},\ }\href {https://doi.org/10.1103/s3vd-brsn} {\bibfield  {journal} {\bibinfo  {journal} {Phys. Rev. Lett.}\ }\textbf {\bibinfo {volume} {135}},\ \bibinfo {pages} {091603} (\bibinfo {year} {2025})}\BibitemShut {NoStop}%
\bibitem [{\citenamefont {Parameswaran}\ \emph {et~al.}(2012)\citenamefont {Parameswaran}, \citenamefont {Kivelson}, \citenamefont {Rezayi}, \citenamefont {Simon}, \citenamefont {Sondhi},\ and\ \citenamefont {Spivak}}]{Kivelson_2012}%
  \BibitemOpen
  \bibfield  {author} {\bibinfo {author} {\bibfnamefont {S.~A.}\ \bibnamefont {Parameswaran}}, \bibinfo {author} {\bibfnamefont {S.~A.}\ \bibnamefont {Kivelson}}, \bibinfo {author} {\bibfnamefont {E.~H.}\ \bibnamefont {Rezayi}}, \bibinfo {author} {\bibfnamefont {S.~H.}\ \bibnamefont {Simon}}, \bibinfo {author} {\bibfnamefont {S.~L.}\ \bibnamefont {Sondhi}},\ and\ \bibinfo {author} {\bibfnamefont {B.~Z.}\ \bibnamefont {Spivak}},\ }\bibfield  {title} {\bibinfo {title} {Typology for quantum {Hall} liquids},\ }\href {https://doi.org/10.1103/PhysRevB.85.241307} {\bibfield  {journal} {\bibinfo  {journal} {Phys. Rev. B}\ }\textbf {\bibinfo {volume} {85}},\ \bibinfo {pages} {241307} (\bibinfo {year} {2012})}\BibitemShut {NoStop}%
\bibitem [{\citenamefont {Deser}\ \emph {et~al.}(2000)\citenamefont {Deser}, \citenamefont {Jackiw},\ and\ \citenamefont {Templeton}}]{Templeton_2000}%
  \BibitemOpen
  \bibfield  {author} {\bibinfo {author} {\bibfnamefont {S.}~\bibnamefont {Deser}}, \bibinfo {author} {\bibfnamefont {R.}~\bibnamefont {Jackiw}},\ and\ \bibinfo {author} {\bibfnamefont {S.}~\bibnamefont {Templeton}},\ }\bibfield  {title} {\bibinfo {title} {Topologically massive gauge theories},\ }\href {https://doi.org/https://doi.org/10.1006/aphy.2000.6013} {\bibfield  {journal} {\bibinfo  {journal} {Ann. Phys.}\ }\textbf {\bibinfo {volume} {281}},\ \bibinfo {pages} {409} (\bibinfo {year} {2000})}\BibitemShut {NoStop}%
\bibitem [{\citenamefont {Gudnason}\ and\ \citenamefont {Speight}(2020)}]{Gudnason_2020}%
  \BibitemOpen
  \bibfield  {author} {\bibinfo {author} {\bibfnamefont {S.~B.}\ \bibnamefont {Gudnason}}\ and\ \bibinfo {author} {\bibfnamefont {J.~M.}\ \bibnamefont {Speight}},\ }\bibfield  {title} {\bibinfo {title} {Realistic classical binding energies in the $\omega$-{Skyrme} model},\ }\href {https://doi.org/10.1007/JHEP07(2020)184} {\bibfield  {journal} {\bibinfo  {journal} {J. High Energ. Phys.}\ }\textbf {\bibinfo {volume} {2020}},\ \bibinfo {pages} {184}}\BibitemShut {NoStop}%
\bibitem [{\citenamefont {Harland}\ \emph {et~al.}(2024)\citenamefont {Harland}, \citenamefont {Leask},\ and\ \citenamefont {Speight}}]{Leask_Harland_2024}%
  \BibitemOpen
  \bibfield  {author} {\bibinfo {author} {\bibfnamefont {D.}~\bibnamefont {Harland}}, \bibinfo {author} {\bibfnamefont {P.}~\bibnamefont {Leask}},\ and\ \bibinfo {author} {\bibfnamefont {M.}~\bibnamefont {Speight}},\ }\bibfield  {title} {\bibinfo {title} {Skyrmion crystals stabilized by $\omega$-mesons},\ }\href {https://doi.org/10.1007/JHEP06(2024)116} {\bibfield  {journal} {\bibinfo  {journal} {J. High Energ. Phys.}\ }\textbf {\bibinfo {volume} {06}},\ \bibinfo {pages} {116}}\BibitemShut {NoStop}%
\bibitem [{\citenamefont {Leask}(2024)}]{Leask_2024}%
  \BibitemOpen
  \bibfield  {author} {\bibinfo {author} {\bibfnamefont {P.}~\bibnamefont {Leask}},\ }\bibfield  {title} {\bibinfo {title} {Baby skyrmion crystals stabilized by vector mesons},\ }\href {https://doi.org/10.1016/j.physletb.2024.138842} {\bibfield  {journal} {\bibinfo  {journal} {Phys. Lett. B}\ }\textbf {\bibinfo {volume} {855}},\ \bibinfo {pages} {138842} (\bibinfo {year} {2024})}\BibitemShut {NoStop}%
\bibitem [{\citenamefont {Nielsen}\ and\ \citenamefont {Olesen}(1973)}]{Nielsen_1973}%
  \BibitemOpen
  \bibfield  {author} {\bibinfo {author} {\bibfnamefont {H.~B.}\ \bibnamefont {Nielsen}}\ and\ \bibinfo {author} {\bibfnamefont {P.}~\bibnamefont {Olesen}},\ }\bibfield  {title} {\bibinfo {title} {Vortex-line models for dual strings},\ }\href {https://doi.org/10.1016/0550-3213(73)90350-7} {\bibfield  {journal} {\bibinfo  {journal} {Nucl. Phys. B}\ }\textbf {\bibinfo {volume} {61}},\ \bibinfo {pages} {45} (\bibinfo {year} {1973})}\BibitemShut {NoStop}%
\bibitem [{\citenamefont {Hindmarsh}\ and\ \citenamefont {Kibble}(1995)}]{Hindmarsh_1995}%
  \BibitemOpen
  \bibfield  {author} {\bibinfo {author} {\bibfnamefont {M.~B.}\ \bibnamefont {Hindmarsh}}\ and\ \bibinfo {author} {\bibfnamefont {T.~W.~B.}\ \bibnamefont {Kibble}},\ }\bibfield  {title} {\bibinfo {title} {Cosmic strings},\ }\href {https://doi.org/10.1088/0034-4885/58/5/001} {\bibfield  {journal} {\bibinfo  {journal} {Rep. Prog. Phys.}\ }\textbf {\bibinfo {volume} {58}},\ \bibinfo {pages} {477} (\bibinfo {year} {1995})}\BibitemShut {NoStop}%
\bibitem [{\citenamefont {Babaev}(2002)}]{Babaev_2002}%
  \BibitemOpen
  \bibfield  {author} {\bibinfo {author} {\bibfnamefont {E.}~\bibnamefont {Babaev}},\ }\bibfield  {title} {\bibinfo {title} {Vortices with fractional flux in two-gap superconductors and in extended {Faddeev} model},\ }\href {https://doi.org/10.1103/PhysRevLett.89.067001} {\bibfield  {journal} {\bibinfo  {journal} {Phys. Rev. Lett.}\ }\textbf {\bibinfo {volume} {89}},\ \bibinfo {pages} {067001} (\bibinfo {year} {2002})}\BibitemShut {NoStop}%
\bibitem [{\citenamefont {Babaev}\ and\ \citenamefont {Speight}(2005)}]{Babaev_Speight_2005}%
  \BibitemOpen
  \bibfield  {author} {\bibinfo {author} {\bibfnamefont {E.}~\bibnamefont {Babaev}}\ and\ \bibinfo {author} {\bibfnamefont {M.}~\bibnamefont {Speight}},\ }\bibfield  {title} {\bibinfo {title} {{Semi-Meissner state and neither type-I nor type-II superconductivity in multicomponent superconductors}},\ }\href {https://doi.org/10.1103/PhysRevB.72.180502} {\bibfield  {journal} {\bibinfo  {journal} {Phys. Rev. B}\ }\textbf {\bibinfo {volume} {72}},\ \bibinfo {pages} {180502} (\bibinfo {year} {2005})}\BibitemShut {NoStop}%
\bibitem [{\citenamefont {Babaev}\ \emph {et~al.}(2010)\citenamefont {Babaev}, \citenamefont {Carlstr\"om},\ and\ \citenamefont {Speight}}]{Babaev_Carlstrom_2010}%
  \BibitemOpen
  \bibfield  {author} {\bibinfo {author} {\bibfnamefont {E.}~\bibnamefont {Babaev}}, \bibinfo {author} {\bibfnamefont {J.}~\bibnamefont {Carlstr\"om}},\ and\ \bibinfo {author} {\bibfnamefont {M.}~\bibnamefont {Speight}},\ }\bibfield  {title} {\bibinfo {title} {{Type-1.5 Superconducting State from an Intrinsic Proximity Effect in Two-Band Superconductors}},\ }\href {https://doi.org/10.1103/PhysRevLett.105.067003} {\bibfield  {journal} {\bibinfo  {journal} {Phys. Rev. Lett.}\ }\textbf {\bibinfo {volume} {105}},\ \bibinfo {pages} {067003} (\bibinfo {year} {2010})}\BibitemShut {NoStop}%
\bibitem [{\citenamefont {Carlstr\"om}\ \emph {et~al.}(2011)\citenamefont {Carlstr\"om}, \citenamefont {Babaev},\ and\ \citenamefont {Speight}}]{Babaev_2011}%
  \BibitemOpen
  \bibfield  {author} {\bibinfo {author} {\bibfnamefont {J.}~\bibnamefont {Carlstr\"om}}, \bibinfo {author} {\bibfnamefont {E.}~\bibnamefont {Babaev}},\ and\ \bibinfo {author} {\bibfnamefont {M.}~\bibnamefont {Speight}},\ }\bibfield  {title} {\bibinfo {title} {Type-1.5 superconductivity in multiband systems: Effects of interband couplings},\ }\href {https://doi.org/10.1103/PhysRevB.83.174509} {\bibfield  {journal} {\bibinfo  {journal} {Phys. Rev. B}\ }\textbf {\bibinfo {volume} {83}},\ \bibinfo {pages} {174509} (\bibinfo {year} {2011})}\BibitemShut {NoStop}%
\bibitem [{\citenamefont {Babaev}\ \emph {et~al.}(2017)\citenamefont {Babaev}, \citenamefont {Carlström}, \citenamefont {Silaev},\ and\ \citenamefont {Speight}}]{Babaev_2017}%
  \BibitemOpen
  \bibfield  {author} {\bibinfo {author} {\bibfnamefont {E.}~\bibnamefont {Babaev}}, \bibinfo {author} {\bibfnamefont {J.}~\bibnamefont {Carlström}}, \bibinfo {author} {\bibfnamefont {M.}~\bibnamefont {Silaev}},\ and\ \bibinfo {author} {\bibfnamefont {J.}~\bibnamefont {Speight}},\ }\bibfield  {title} {\bibinfo {title} {Type-1.5 superconductivity in multicomponent systems},\ }\href {https://doi.org/10.1016/j.physc.2016.08.003} {\bibfield  {journal} {\bibinfo  {journal} {Physica C Supercond.}\ }\textbf {\bibinfo {volume} {533}},\ \bibinfo {pages} {20} (\bibinfo {year} {2017})}\BibitemShut {NoStop}%
\bibitem [{\citenamefont {Samoilenka}\ and\ \citenamefont {Babaev}(2020)}]{Samoilenka_2020}%
  \BibitemOpen
  \bibfield  {author} {\bibinfo {author} {\bibfnamefont {A.}~\bibnamefont {Samoilenka}}\ and\ \bibinfo {author} {\bibfnamefont {E.}~\bibnamefont {Babaev}},\ }\bibfield  {title} {\bibinfo {title} {Spiral magnetic field and bound states of vortices in noncentrosymmetric superconductors},\ }\href {https://doi.org/10.1103/PhysRevB.102.184517} {\bibfield  {journal} {\bibinfo  {journal} {Phys. Rev. B}\ }\textbf {\bibinfo {volume} {102}},\ \bibinfo {pages} {184517} (\bibinfo {year} {2020})}\BibitemShut {NoStop}%
\bibitem [{\citenamefont {Garaud}\ \emph {et~al.}(2020)\citenamefont {Garaud}, \citenamefont {Chernodub},\ and\ \citenamefont {Kharzeev}}]{Garaud_2020}%
  \BibitemOpen
  \bibfield  {author} {\bibinfo {author} {\bibfnamefont {J.}~\bibnamefont {Garaud}}, \bibinfo {author} {\bibfnamefont {M.~N.}\ \bibnamefont {Chernodub}},\ and\ \bibinfo {author} {\bibfnamefont {D.~E.}\ \bibnamefont {Kharzeev}},\ }\bibfield  {title} {\bibinfo {title} {Vortices with magnetic field inversion in noncentrosymmetric superconductors},\ }\href {https://doi.org/10.1103/PhysRevB.102.184516} {\bibfield  {journal} {\bibinfo  {journal} {Phys. Rev. B}\ }\textbf {\bibinfo {volume} {102}},\ \bibinfo {pages} {184516} (\bibinfo {year} {2020})}\BibitemShut {NoStop}%
\bibitem [{\citenamefont {Mu\~noz de~las Heras}\ \emph {et~al.}(2020)\citenamefont {Mu\~noz de~las Heras}, \citenamefont {Macaluso},\ and\ \citenamefont {Carusotto}}]{Heras_2020}%
  \BibitemOpen
  \bibfield  {author} {\bibinfo {author} {\bibfnamefont {A.}~\bibnamefont {Mu\~noz de~las Heras}}, \bibinfo {author} {\bibfnamefont {E.}~\bibnamefont {Macaluso}},\ and\ \bibinfo {author} {\bibfnamefont {I.}~\bibnamefont {Carusotto}},\ }\bibfield  {title} {\bibinfo {title} {Anyonic molecules in atomic fractional quantum hall liquids: A quantitative probe of fractional charge and anyonic statistics},\ }\href {https://doi.org/10.1103/PhysRevX.10.041058} {\bibfield  {journal} {\bibinfo  {journal} {Phys. Rev. X}\ }\textbf {\bibinfo {volume} {10}},\ \bibinfo {pages} {041058} (\bibinfo {year} {2020})}\BibitemShut {NoStop}%
\bibitem [{\citenamefont {Shi}\ and\ \citenamefont {Senthil}(2025)}]{Shi_2025}%
  \BibitemOpen
  \bibfield  {author} {\bibinfo {author} {\bibfnamefont {Z.~D.}\ \bibnamefont {Shi}}\ and\ \bibinfo {author} {\bibfnamefont {T.}~\bibnamefont {Senthil}},\ }\bibfield  {title} {\bibinfo {title} {Doping a fractional quantum anomalous hall insulator},\ }\href {https://doi.org/10.1103/kcm5-hx56} {\bibfield  {journal} {\bibinfo  {journal} {Phys. Rev. X}\ }\textbf {\bibinfo {volume} {15}},\ \bibinfo {pages} {031069} (\bibinfo {year} {2025})}\BibitemShut {NoStop}%
\bibitem [{\citenamefont {Kuhlenkamp}\ \emph {et~al.}(2024)\citenamefont {Kuhlenkamp}, \citenamefont {Kadow}, \citenamefont {Imamoğlu},\ and\ \citenamefont {Knap}}]{Kuhlenkamp_2024}%
  \BibitemOpen
  \bibfield  {author} {\bibinfo {author} {\bibfnamefont {C.}~\bibnamefont {Kuhlenkamp}}, \bibinfo {author} {\bibfnamefont {W.}~\bibnamefont {Kadow}}, \bibinfo {author} {\bibfnamefont {A.}~\bibnamefont {Imamoğlu}},\ and\ \bibinfo {author} {\bibfnamefont {M.}~\bibnamefont {Knap}},\ }\bibfield  {title} {\bibinfo {title} {Chiral pseudospin liquids in moir\'e heterostructures},\ }\href {https://doi.org/10.1103/PhysRevX.14.021013} {\bibfield  {journal} {\bibinfo  {journal} {Phys. Rev. X}\ }\textbf {\bibinfo {volume} {14}},\ \bibinfo {pages} {021013} (\bibinfo {year} {2024})}\BibitemShut {NoStop}%
\bibitem [{\citenamefont {Speight}\ and\ \citenamefont {Winyard}(2025{\natexlab{a}})}]{Speight_Winyard_2025}%
  \BibitemOpen
  \bibfield  {author} {\bibinfo {author} {\bibfnamefont {M.}~\bibnamefont {Speight}}\ and\ \bibinfo {author} {\bibfnamefont {T.}~\bibnamefont {Winyard}},\ }\bibfield  {title} {\bibinfo {title} {Vortex lattices and critical fields in anisotropic superconductors},\ }\href {https://doi.org/10.1088/1751-8121/adb7a7} {\bibfield  {journal} {\bibinfo  {journal} {J. Phys. A: Math. Theor.}\ }\textbf {\bibinfo {volume} {58}},\ \bibinfo {pages} {095203} (\bibinfo {year} {2025}{\natexlab{a}})}\BibitemShut {NoStop}%
\bibitem [{\citenamefont {Bazeia}\ \emph {et~al.}(2024)\citenamefont {Bazeia}, \citenamefont {Campos},\ and\ \citenamefont {Mohammadi}}]{Bazeia_2024}%
  \BibitemOpen
  \bibfield  {author} {\bibinfo {author} {\bibfnamefont {D.}~\bibnamefont {Bazeia}}, \bibinfo {author} {\bibfnamefont {J.~G.~F.}\ \bibnamefont {Campos}},\ and\ \bibinfo {author} {\bibfnamefont {A.}~\bibnamefont {Mohammadi}},\ }\bibfield  {title} {\bibinfo {title} {{Abelian Chern-Simons vortices in the presence of magnetic impurities}},\ }\href {https://doi.org/10.1007/JHEP12(2024)108} {\bibfield  {journal} {\bibinfo  {journal} {J. High Energ. Phys.}\ }\textbf {\bibinfo {volume} {12}},\ \bibinfo {pages} {108}}\BibitemShut {NoStop}%
\bibitem [{\citenamefont {Speight}\ and\ \citenamefont {Winyard}(2025{\natexlab{b}})}]{Winyard_2025}%
  \BibitemOpen
  \bibfield  {author} {\bibinfo {author} {\bibfnamefont {M.}~\bibnamefont {Speight}}\ and\ \bibinfo {author} {\bibfnamefont {T.}~\bibnamefont {Winyard}},\ }\bibfield  {title} {\bibinfo {title} {Short-range intervortex forces},\ }\href {https://doi.org/10.1103/2m3b-gr79} {\bibfield  {journal} {\bibinfo  {journal} {Phys. Rev. D}\ }\textbf {\bibinfo {volume} {112}},\ \bibinfo {pages} {055024} (\bibinfo {year} {2025}{\natexlab{b}})}\BibitemShut {NoStop}%
\bibitem [{\citenamefont {Barkman}\ \emph {et~al.}(2020)\citenamefont {Barkman}, \citenamefont {Samoilenka}, \citenamefont {Winyard},\ and\ \citenamefont {Babaev}}]{Barkman_2020}%
  \BibitemOpen
  \bibfield  {author} {\bibinfo {author} {\bibfnamefont {M.}~\bibnamefont {Barkman}}, \bibinfo {author} {\bibfnamefont {A.}~\bibnamefont {Samoilenka}}, \bibinfo {author} {\bibfnamefont {T.}~\bibnamefont {Winyard}},\ and\ \bibinfo {author} {\bibfnamefont {E.}~\bibnamefont {Babaev}},\ }\bibfield  {title} {\bibinfo {title} {Ring solitons and soliton sacks in imbalanced fermionic systems},\ }\href {https://doi.org/10.1103/PhysRevResearch.2.043282} {\bibfield  {journal} {\bibinfo  {journal} {Phys. Rev. Res.}\ }\textbf {\bibinfo {volume} {2}},\ \bibinfo {pages} {043282} (\bibinfo {year} {2020})}\BibitemShut {NoStop}%
\bibitem [{\citenamefont {Paul}\ and\ \citenamefont {Khare}(1986)}]{Paul_1986}%
  \BibitemOpen
  \bibfield  {author} {\bibinfo {author} {\bibfnamefont {S.~K.}\ \bibnamefont {Paul}}\ and\ \bibinfo {author} {\bibfnamefont {A.}~\bibnamefont {Khare}},\ }\bibfield  {title} {\bibinfo {title} {{Charged vortices in an abelian Higgs model with Chern-Simons term}},\ }\href {https://doi.org/10.1016/0370-2693(86)91028-2} {\bibfield  {journal} {\bibinfo  {journal} {Phys. Lett. B}\ }\textbf {\bibinfo {volume} {174}},\ \bibinfo {pages} {420} (\bibinfo {year} {1986})}\BibitemShut {NoStop}%
\bibitem [{\citenamefont {Pisarski}\ and\ \citenamefont {Rao}(1985)}]{Pisarski_1985}%
  \BibitemOpen
  \bibfield  {author} {\bibinfo {author} {\bibfnamefont {R.~D.}\ \bibnamefont {Pisarski}}\ and\ \bibinfo {author} {\bibfnamefont {S.}~\bibnamefont {Rao}},\ }\bibfield  {title} {\bibinfo {title} {Topologically massive chromodynamics in the perturbative regime},\ }\href {https://doi.org/10.1103/PhysRevD.32.2081} {\bibfield  {journal} {\bibinfo  {journal} {Phys. Rev. D}\ }\textbf {\bibinfo {volume} {32}},\ \bibinfo {pages} {2081} (\bibinfo {year} {1985})}\BibitemShut {NoStop}%
\bibitem [{\citenamefont {Speight}(1997)}]{Speight_1997}%
  \BibitemOpen
  \bibfield  {author} {\bibinfo {author} {\bibfnamefont {J.~M.}\ \bibnamefont {Speight}},\ }\bibfield  {title} {\bibinfo {title} {Static intervortex forces},\ }\href {https://doi.org/10.1103/PhysRevD.55.3830} {\bibfield  {journal} {\bibinfo  {journal} {Phys. Rev. D}\ }\textbf {\bibinfo {volume} {55}},\ \bibinfo {pages} {3830} (\bibinfo {year} {1997})}\BibitemShut {NoStop}%
\bibitem [{\citenamefont {Manton}\ and\ \citenamefont {Speight}(2003)}]{Manton_Speight_2003}%
  \BibitemOpen
  \bibfield  {author} {\bibinfo {author} {\bibfnamefont {N.~S.}\ \bibnamefont {Manton}}\ and\ \bibinfo {author} {\bibfnamefont {J.~M.}\ \bibnamefont {Speight}},\ }\bibfield  {title} {\bibinfo {title} {Asymptotic interactions of critically coupled vortices},\ }\href {https://doi.org/10.1007/s00220-003-0842-4} {\bibfield  {journal} {\bibinfo  {journal} {Commun. Math. Phys.}\ }\textbf {\bibinfo {volume} {236}},\ \bibinfo {pages} {535} (\bibinfo {year} {2003})}\BibitemShut {NoStop}%
\bibitem [{\citenamefont {Speight}\ and\ \citenamefont {Winyard}(2021)}]{Speight_2021}%
  \BibitemOpen
  \bibfield  {author} {\bibinfo {author} {\bibfnamefont {M.}~\bibnamefont {Speight}}\ and\ \bibinfo {author} {\bibfnamefont {T.}~\bibnamefont {Winyard}},\ }\bibfield  {title} {\bibinfo {title} {Intervortex forces in competing-order superconductors},\ }\href {https://doi.org/10.1103/PhysRevB.103.014514} {\bibfield  {journal} {\bibinfo  {journal} {Phys. Rev. B}\ }\textbf {\bibinfo {volume} {103}},\ \bibinfo {pages} {014514} (\bibinfo {year} {2021})}\BibitemShut {NoStop}%
\bibitem [{\citenamefont {Bettencourt}\ and\ \citenamefont {Rivers}(1995)}]{Bettencourt_1995}%
  \BibitemOpen
  \bibfield  {author} {\bibinfo {author} {\bibfnamefont {L.~M.~A.}\ \bibnamefont {Bettencourt}}\ and\ \bibinfo {author} {\bibfnamefont {R.~J.}\ \bibnamefont {Rivers}},\ }\bibfield  {title} {\bibinfo {title} {{Interactions between U(1) cosmic strings: An analytical study}},\ }\href {https://doi.org/10.1103/PhysRevD.51.1842} {\bibfield  {journal} {\bibinfo  {journal} {Phys. Rev. D}\ }\textbf {\bibinfo {volume} {51}},\ \bibinfo {pages} {1842} (\bibinfo {year} {1995})}\BibitemShut {NoStop}%
\bibitem [{\citenamefont {Fujikura}\ \emph {et~al.}(2023)\citenamefont {Fujikura}, \citenamefont {Li},\ and\ \citenamefont {Yamaguchi}}]{Fujikura_2023}%
  \BibitemOpen
  \bibfield  {author} {\bibinfo {author} {\bibfnamefont {K.}~\bibnamefont {Fujikura}}, \bibinfo {author} {\bibfnamefont {S.}~\bibnamefont {Li}},\ and\ \bibinfo {author} {\bibfnamefont {M.}~\bibnamefont {Yamaguchi}},\ }\bibfield  {title} {\bibinfo {title} {Interactions between several types of cosmic strings},\ }\href {https://doi.org/10.1007/JHEP12(2023)115} {\bibfield  {journal} {\bibinfo  {journal} {J. High Energ. Phys.}\ }\textbf {\bibinfo {volume} {12}},\ \bibinfo {pages} {115}}\BibitemShut {NoStop}%
\end{thebibliography}%


\appendix


\subsection{Single-vortex anyon tail}
\label{subsec: Single-vortex anyon tail}

Let the vacuum (ground state) have modulus $|\psi|=m>0$, and define the Proca mass $m_A=qm$, with $q$ the gauge charge.
We work in the Coulomb gauge $\partial_i A_i=0$ and linearize about
\begin{equation}
\label{eq: Vacuum}
    \psi = m + \phi(r), \quad A_\mu = 0 +a_\mu,
\end{equation}
with $|\phi|\ll m$ and the unitary gauge (zero phase) for the Higgs field mode.
In general, the Taylor expansion of the potential $V(|\psi|)$ about the ground state $|\psi|=m$ is $V(|\psi|) = V(m) + \frac{1}{2}\phi^2 V''(m) + O(\phi^3)$.
This gives the Higgs field a mass $m_H=\sqrt{V''(m)}=m\sqrt{\lambda}$.
In terms of the gauge field mode $a_\mu$, the associated magnetic field is $B=\varepsilon_{ij}\partial_i a_j$ and the electric field is $E_i=-\partial_i a_0$.
Then, the static energy functional, linearized about the vacuum \eqref{eq: Vacuum}, is found to be given by
\begin{widetext}
    \begin{align}
    \label{eq: Linearized energy}
        E_{\textup{lin}} = \, & \frac{1}{2} \int_{\mathbb{R}^2} \textup{d}^2x \left[ \phi  \left( -\nabla^2 + m_H^2 \right) \phi \right]  + \frac{1}{2} \int_{\mathbb{R}^2} \textup{d}^2x
        \begin{bmatrix}
            B & a_0
        \end{bmatrix}
        \begin{bmatrix}
            \left( -\nabla^2 + m_A^2 \right) & -\kappa\nabla^2 \\
            \kappa & \left( -\nabla^2 + m_A^2 \right)
        \end{bmatrix}
        \begin{bmatrix}
            B \\
            a_0
        \end{bmatrix}.
    \end{align}
\end{widetext}

To linear order, the Gauss constraint \eqref{eq: Gauss law} and the static Ginzburg--Landau equation \eqref{eq: Field equations} reduce to
\begin{subequations}
    \begin{align}
        \left( -\nabla^2 + m_H^2 \right) \phi = \, & 0, \label{eq: Linearized psi} \\
        \left( -\nabla^2 + m_A^2 \right) a_i = \, & \kappa \epsilon_{ij} \partial_j a_0, \label{eq: Linearized B} \\
        \left( -\nabla^2 + m_A^2 \right) a_0 = \, & -\kappa B. \label{eq: Linearized E}
    \end{align}
\end{subequations}
It can be seen that the Higgs-amplitude mode $\phi$ decouples at quadratic order, giving a static Klein-Gordon equation.
Let us take the curl of \eqref{eq: Linearized B}, which yields
\begin{equation}
\label{eq: Curl of linearized B}
    (-\nabla^2 + m_A^2)B = \kappa \nabla^2 a_0, 
\end{equation}
and apply the Laplace operator to the linearized Gauss' law \eqref{eq: Linearized E}, giving us
\begin{equation}
\label{eq: Laplacian of linearized E}
    (\nabla^2 + m_A^2) \nabla^2a_0 = -\kappa \nabla^2 B.
\end{equation}
Then, applying the operator $(\nabla^2 + m_A^2)$ to \eqref{eq: Curl of linearized B} and using the relation \eqref{eq: Laplacian of linearized E}, we obtain a scalar decoupled fourth order Helmholtz like equation for the magnetic field,
\begin{equation}
    \left[ \left( -\nabla^2 + m_A^2 \right)^2 + \kappa^2\nabla^2 \right] B=0.
\end{equation}
This linearized equation takes the same form as the linearized field equation for the order parameter in superfluids with fermionic imbalance \cite{Barkman_2020}.
We can factorize this operator into complex-conjugate eigenmodes,
\begin{align}
    \Delta_{\kappa} \equiv \, & \left[ \nabla^4 - \left(2m_A^2-\kappa^2\right)\nabla^2 + m_A^4 \right] \nonumber \\
    = \, & \left( \nabla^2 - m_+^2 \right)\left( \nabla^2 - m_-^2 \right),
\end{align}
where the two complex-conjugate gauge eigenmodes are
\begin{equation}
    m_\pm = \alpha \pm i \beta, \quad \alpha = \sqrt{m_A^2-\frac{\kappa^2}{4}}, \quad \beta = \frac{\kappa}{2},
\end{equation}
and they satisfy the following relations,
\begin{equation}
    m_+ m_- = m_A^2, \quad m_+^2 + m_-^2 = 2m_A^2 - \kappa^2.
\end{equation}
We also get the exact same operator acting on the electric field, that is $\Delta_{\kappa} \vec{E}=\vec{0}$.
Hence, the complex-conjugate eigenmode pair $m_\pm$ are the screening masses of the electric and magnetic fields.
To see this, we simply look for asymptotic homogeneous modes with $B(r)\propto e^{-mr}$.
Equivalently, we set $\nabla^2\mapsto m^2$ in the operator $\Delta_{\kappa}$, which yields a characteristic equation whose solutions are exactly $m=m_\pm$.
Thus, instead of two independent gauge coherence lengths, we have one gauge length plus an oscillation scale set by $\beta$, where the two gauge modes share the same decay rate $\alpha$.
The corresponding common gauge coherence (screening) length and oscillation length are
\begin{equation}
    \lambda_{\textup{gauge}} = \frac{1}{\alpha} = \frac{1}{\sqrt{m_A^2-\kappa^2/4}}, \quad \lambda_{\textup{osc}}=\frac{2\pi}{\beta}=\frac{4\pi}{\kappa}.
\end{equation}
So, the difference between the magnetic field $B$ and the electric field $\vec{E}$ asymptotically is an oscillatory phase, not the decay rate.

This disagrees with the result of Paul--Khare \cite{Paul_1986}, where they claim to find a pair of real penetration depths $\lambda_\pm=1/M_\pm$, with real static screening masses $M_\pm=\sqrt{m_A^2+\kappa^2/4}\,\pm\kappa/2$.
The static screening masses they obtain is associated to the operator $[(-\nabla^2 + m_A^2)^2 - \kappa^2\nabla^2]$, which has the wrong sign for the linearized static theory.
These masses were originally obtained by Pisarski--Rao \cite{Pisarski_1985}.
However, a subtle distinction is that the masses Pisarski--Rao determined are the propagator pole (dynamical gauge) masses $M_\pm$, not the static screening masses $m_\pm$.
They are related to the static screening masses by a Wick rotation $\kappa \mapsto i\kappa$.
The penetration depths are related to the static screening masses, not the propagator pole masses.
In the simplest of cases (such as the Ginzburg--Landau model with $\kappa=0$), the propagator pole and static screening masses coincide, however this is not true when the Chern--Simons term \eqref{eq: Chern--Simons term} is included.

Let us think about this from another perspective.
Since the magnetic flux $\Phi$ and the electric charge $Q_m$ are related via the relation \eqref{eq: Noether electric charge}, the electric and magnetic sectors are not independent.
That is, a fluctuation in the magnetic field $B$ must be accompanied by a fluctuation in the potential $A_0$ and, hence, the electric field $\vec{E}=-\vec{\nabla}A_0$.
This is why the linearized spectrum does not split into two independent real gauge masses.
Instead we get a single decay rate and an oscillation frequency, i.e. a damped oscillatory profile for both the magnetic field and the electric field.
So the equality of the two gauge coherence lengths is exactly the linearized reflection of the magnetic flux-charge constraint.

To summarize, we find that the single vortex anyon far tails (asymptotic form) take the form
\begin{subequations}
\label{eq: Asymptotic forms}
    \begin{align}
        \phi(r) = \, & c_HK_0(m_H r), \label{eq: Higgs tail}\\
        B(r) = \, & c_B K_0(m_+ r) + c_B^\ast K_0(m_- r), \label{eq: B tail}\\
        a_0(r) = \, & c_E K_0(m_+ r) + c_E^\ast K_0(m_- r), \label{eq: E tail}
    \end{align}
\end{subequations}
where $c_H\in\mathbb{R}$ and $c_{B,E}\in\mathbb{C}$.
It is clear that the magnetic and electric field are manifestly real, since $m_-=m_+^\ast$ and $K_0$ is analytic.


\subsection{Long-range interactions}
\label{subsec: Long-range interactions}

Now that we have determined the single-vortex far field form, we proceed to determine the long-range interaction energy following the point-particle method \cite{Speight_1997,Manton_Speight_2003,Speight_2021}.
This is done in two parts, we (i) add linear sources to the linearized energy \eqref{eq: Linearized energy} such that the solutions of the corresponding field equations are exactly the single-vortex far fields, and (ii) compute the interaction energy from the on-shell cross term in the linearization.

Let us introduce sources $S_H$, $S_B$, and $S_E$ that couple linearly to $\phi$, $B$, and $a_0$, respectively.
This gives us additional terms in the linearized energy,
\begin{equation}
    E_{\textup{source}} = -\int_{\mathbb{R}^2}\textup{d}^2x \left[ S_H \phi + S_B B + S_E a_0 \right].
\end{equation}
The sourced linear equations, associated to $E=E_{\textup{lin}}+E_{\textup{source}}$, are determined to be
\begin{subequations}
    \begin{align}
        (-\nabla^2 + m_H^2) \phi = \, & S_H, \label{eq:hlin}\\
        (-\nabla^2 + m_A^2) B - \kappa \nabla^2 a_0 = \, & S_B, \label{eq:Blinsrc}\\
        (-\nabla^2 + m_A^2) a_0 - \kappa B = \, & S_E. \label{eq:A0linsrc}
    \end{align}
\end{subequations}
As we did previously, we can eliminate the potential mode $a_0$ giving a driven fourth-order equation for the magnetic field $B$,
\begin{equation}
  \Delta_{\kappa} B = (-\nabla^2+m_A^2)S_B + \kappa\nabla^2\,S_E.
  \label{eq: Driven}
\end{equation}
The homogeneous part of this ($\Delta_{\kappa}B=0$) retains the far field asymptotics.

The Green's function for the static Klein-Gordon equation in 2D is the Bessel function $K_0$, that is, it satisfies
\begin{equation}
    \left( -\nabla^2 + \mu^2 \right) K_0(\mu r) = 2\pi\delta(r).
\end{equation}
Therefore, we see that the appropriate source for the Higgs amplitude mode is
\begin{equation}
    S_H(r) = c_H(-\nabla^2 + m_H^2)K_0(m_H r) = 2\pi c_H \delta(r).
\label{eq: Higgs source}
\end{equation}
In order to realize exactly the far field form \eqref{eq: B tail} of the magnetic field with real fields, it is convenient to choose the electric source to be $S_E(r)=0$ and then we find the magnetic source to be given by
\begin{equation}
    S_B(r)= 2\pi\left[
    c_B\frac{(-\nabla^2 + m_-^2)}{(-\nabla^2 + m_A^2)}
    + c_B^\ast\frac{(-\nabla^2 + m_+^2)}{(-\nabla^2 + m_A^2)}
    \right]\delta(r).
\label{eq: Magnetic field source}
\end{equation}
We note that the $(-\nabla^2 + m_A^2)^{-1}$ operator acting on the Dirac $\delta$-function is just the Yukawa kernel with mass $M$.
Operationally, \eqref{eq: Magnetic field source} is a linear combination of $\delta$ and $\nabla^2\delta$.
Substituting the magnetic field source \eqref{eq: Magnetic field source} into \eqref{eq: Driven} yields
\begin{equation}
    \Delta_{\kappa} B = 2\pi \left[c_B(-\nabla^2+m_-^{\,2}) + c_B^\ast(-\nabla^2+m_+^{\,2})\right]\delta(r),
\end{equation}
whose solution is precisely \eqref{eq: B tail}.
We can then determine the electric potential $a_0$ from Gauss' law \eqref{eq: Gauss law},
\begin{equation}
    a_0(r) = -\kappa\left[ \frac{c_B}{m_A^2 - m_+^2}K_0(m_+ r) + \frac{c_B^\ast}{m_A^2-m_-^2} K_0(m_- r) \right],
\label{eq:A0tail}
\end{equation}
from which we can identify
\begin{equation}
    c_E = -\kappa\frac{c_B}{m_A^2 - m_+^2}.
\end{equation}

With the linear sources in-hand, we turn to computing the asymptotic interaction energy of well-separated vortices.
Let us consider two identical vortices placed at $\vec{X}_1$ and $\vec{X}_2$, each with sources \eqref{eq: Higgs source}, \eqref{eq: Magnetic field source} translated to its center.
Let us define the separation distance $R=|\vec{X}_1-\vec{X}_2|$ and vortex centres $r_i=|\vec{x}-\vec{X}_i|$.
The interaction energy between well-separated sources comes from the cross-terms in the linearization.
That is,
\begin{align}
    V_{\textup{int}}(R) = \, & -\int_{\mathbb{R}^2}\textup{d}^2\vec{x} \left[ S_H^{(1)} \phi^{(2)} +S_B^{(1)} B^{(2)} \right] \nonumber \\
    = \, &-\int_{\mathbb{R}^2}\textup{d}^2\vec{x} \left[ S_H(r_1) \phi(r_2) +S_B(r_1) B(r_2) \right] .
\end{align}

The Higgs channel gives the usual attractive interaction contribution \cite{Bettencourt_1995}
\begin{align}
    V_H(R) = \, & -\int_{\mathbb{R}^2}\textup{d}^2\vec{x} \left[ 2\pi c_H^2 \delta(r_1) K_0(r_2) \right] \nonumber \\
    = \, & -2\pi c_H^2 K_0(m_H R),
\end{align}
and the gauge channel yields
\begin{align}
    V_B(R) = \, & 2\pi\int_{\mathbb{R}^2}\textup{d}^2\vec{x} \delta(r_1) \left[ c_B^2 K_0(m_+ r_2) + (c_B^\ast)^2 K_0(m_- r_2) \right] \nonumber \\
    = \, & 2\pi \left[ c_B^2 K_0(m_+ R) + (c_B^\ast)^2 K_0(m_- R) \right] \nonumber \\
    = \, & 4\pi \, \textup{Re}\left[ c_B^2 K_0(m_+ R) \right].
\end{align}
Let us write the screening mass in the form $m_+=\alpha+i\beta=\rho e^{i\theta}$ with $\rho=\sqrt{\alpha^2+\beta^2}=m_A$ and $\theta=\arg(m_+)$.
Then, a power series expansion of the Bessel function $K_0$ at large separation $R$ yields
\begin{equation}
    K_0(m_+ R) \sim \sqrt{\frac{\pi}{2m_A R}} \, e^{-\alpha R} \, e^{-i(\beta R+\theta/2)}.
\end{equation}
Hence, the gauge channel interaction energy becomes
\begin{equation}
    V_B(R) = 4\pi |c_B|^2 \sqrt{\frac{\pi}{2m_A R}} \, e^{-\alpha R} \cos\left( \beta R - \gamma \right),
\end{equation}
where $\gamma=\theta/2-2\vartheta$ and $\vartheta=\arg(c_B)$.
So, we see that the long-range oscillation behavior of the gauge modes leads to more interesting long-range interactions.

Putting this all together, we find that the interaction energy of a pair of separated vortex anyons is determined by
\begin{align}
    V_{\textup{int}}(R) = \, & 2\pi |c_B|^2 \sqrt{\frac{2\pi}{m_A R}} \, e^{-\alpha R} \cos\left( \beta R - \gamma \right)  \nonumber \\
    \, & - 2\pi c_H^2 K_0(m_H R).
\end{align}


\subsection{Ginzburg--Landau limit $\kappa\to 0$}
\label{subsec: Ginzburg--Landau limit}

As a consistency check, we want to ensure that we recover the Ginzburg--Landau model in the limit $\kappa\to 0$.
First of all, we observe that the decay rate becomes $\lim_{\kappa\rightarrow 0}\alpha=m_A=qm$ and the oscillatory behavior vanishes, $\lim_{\kappa\rightarrow 0}\beta = 0$.
Therefore, the magnetic penetration depth is recovered,
\begin{equation}
    \lim_{\kappa\rightarrow 0} \lambda_{\textup{gauge}} = \frac{1}{\alpha} = \frac{1}{m_A},
\end{equation}
and the complex-conjugate screening masses tend to the single real-valued Proca mass,
\begin{equation}
    \lim_{\kappa\rightarrow 0} m_\pm = \alpha = m_A.
\end{equation}
Furthermore, we recover the long-range interaction energy of the Ginzburg--Landau model \cite{Bettencourt_1995,Fujikura_2023},
\begin{equation}
    \lim_{\kappa\rightarrow 0} V_{\textup{int}}(R) = 2\pi \left[ c_A^2 K_0(m_AR) - c_H^2 K_0(m_H R) \right],
\end{equation}
where $c_A^2=2|c_B|^2$.

\end{document}